# Two-loop relation between the bare lattice coupling and the $\overline{\mathrm{MS}}$ coupling in pure SU($N$) gauge theories

Martin Lüscher

Deutsches Elektronen-Synchrotron DESY
Notkestrasse 85, D-22603 Hamburg, Germany

Peter Weisz

Max-Planck-Institut für Physik
Föhringer Ring 6, D-80805 München, Germany

**Abstract**

We report the result of a computation of the relation between the renormalized coupling in the $\overline{\mathrm{MS}}$ scheme of dimensional regularization and the bare coupling in the standard lattice formulation of the SU($N$) Yang-Mills theory to two-loop order of perturbation theory and discuss some of its implications.

1. The possibility to compute the running coupling in asymptotically free theories through numerical simulations of the corresponding lattice theories has recently attracted much attention and a lot of progress has been made in this field [1–16]. Different computational strategies are being pursued and several non-perturbative definitions of the coupling have been considered. In the pure SU(2) and SU(3) gauge theories the so far most precise and complete results have been obtained for a coupling $\alpha_{\rm SF}(q)$ related to the Schrödinger functional of the theory [2–5,8,9].

At sufficiently large momenta $q$ any two running couplings can be expanded in powers of each other and the choices which one has made then become largely irrelevant. In particular, through the series

$$\alpha_{\overline{\rm MS}}(sq) = \alpha_{\rm SF}(q) + c_1(s)\,\alpha_{\rm SF}(q)^2 + c_2(s)\,\alpha_{\rm SF}(q)^3 + \ldots \tag{1.1}$$

(where $s$ is an adjustable scale factor), one can convert from the SF scheme to the $\overline{\rm MS}$ scheme of dimensional regularization [17,18], the currently most widely used scheme for the analysis of experimental data in high-energy physics.

The one-loop coefficient $c_1(s)$ in the Yang-Mills theory with gauge group SU(2) and SU(3) has been calculated in refs.[2,5]. Considering the precision which can presently be reached in the numerical simulations, it is now desirable to extend the expansion (1.1) to two-loop order. In principle this could be done by evaluating the Schrödinger functional to this order using dimensional regularization. Inspection shows, however, that such a calculation would be technically delicate and computationally non-trivial. It seemed much safer to us to first expand the Schrödinger functional in the lattice theory in powers of the bare coupling $\alpha_0$ and to combine the result with the series

$$\alpha_{\overline{\rm MS}}(s/a) = \alpha_0 + d_1(s)\,\alpha_0^2 + d_2(s)\,\alpha_0^3 + \ldots, \tag{1.2}$$

which one derives on an infinite lattice with spacing $a$.

For the case of the SU(2) theory the expansion of the lattice Schrödinger functional to two-loop order has just been completed by Narayanan and Wolff [19,20]. For gauge group SU($N$) the one-loop coefficient $d_1(s)$ appearing above has been computed nearly 15 years ago by A. and P. Hasenfratz [21] and an estimate for $d_2(s)$ has been obtained by Ellis and Martinelli [22–24] on the basis of "tadpole dominance" (as first proposed by Parisi [25]). We have now performed a complete calculation of $d_2(s)$ in the SU($N$) theory and discuss our result in this letter. The details of the computation will be published elsewhere.

2. We consider the standard SU($N$) lattice gauge theory on a four-dimensional hypercubic lattice with spacing $a$ and Wilson action. The bare gauge coupling is denoted by $g_0$ and $\alpha_0 = g_0^2/4\pi$. The Feynman rules in this theory can be deduced following well established lines (see e.g. ref.[26]). To determine the relation between $\alpha_0$ and $\alpha_{\overline{\rm MS}}$, a set of correlations functions must be worked out to the desired order of perturbation theory and matched with the corresponding amplitudes which one obtains in the continuum theory using dimensional regularization.

The analytical expressions for the ghost and gluon vertices are quite complicated on the lattice. As a result one tends to end up with Feynman integrands that easily fill several pages of paper even if all external Lorentz indices are contracted. Moreover, the Feynman parameter representation and other widely used tools to evaluate Feynman diagrams are of little help in the lattice theory, because the propagators are not simple quadratic functions in the loop and external momenta. It is, therefore, of crucial importance to arrange the computation in a most economical way.

With this in mind we have decided to make use of the background field technique [27,28], which has previously been applied to compute the one-loop coefficient $d_1(s)$ [29]. The idea of the method is to introduce a classical background gauge field and to couple it to the quantum fields in such a way that the system has extended symmetry properties. An important consequence of these symmetries is that the renormalization of the theory does not require any further counterterms besides those already required in the absence of the background field [28]. The background field should thus be regarded as an additional source field which allows one to probe the theory in a particularly interesting way.

For our purposes the most suitable quantity to consider is the background field propagator, i.e. the second derivative of the effective action with respect to the background field at vanishing sources. To determine the matching of the gauge coupling in different schemes it is in fact sufficient to work out this amplitude to the desired order of perturbation theory. In addition the gauge field propagator must be calculated to lower orders to control the renormalization of the gauge fixing parameter. At two-loop order the computation then amounts to the evaluation of 31 propagator type Feynman diagrams on the lattice and a smaller number of diagrams in the continuum theory. Note that we do not set the external momentum to zero (in contrast to ref.[29]). The Feynman integrals are, therefore, perfectly well-defined and no intermediate infrared regularization is required.

To evaluate the lattice Feynman integrals in the continuum limit, we have first extracted the dependence on the external momentum analytically and then computed the remaining parts of the integrals numerically through a newly developed position space technique. We expect this method to be more widely applicable and shall describe it in a separate publication. With little computer time the integrals which occurred could in this way be calculated to a numerical precision better than 7 significant decimal places.

To guarantee the correctness of our results all lattice diagrams were calculated independently by each author, using different sets of algebraic and numerical programs on different computer systems. We checked algebraically that the quadratically divergent parts cancel in the sum of all diagrams and that the coefficients of the logarithmically divergent terms add up to the value expected from the Callan-Symanzik $\beta$–function. The background field propagator in the $\overline{\mathrm{MS}}$ scheme, which is also needed in our computations, has previously been worked out to two-loop order by Ellis [24]. The formula quoted by Ellis completely agrees with our results.

The outcome of our calculations is summarized by

$$d_1(s) = -\frac{11N}{6\pi}\ln(s) - \frac{\pi}{2N} + k_1 N, \tag{2.1}$$

$$d_2(s) = d_1(s)^2 - \frac{17N^2}{12\pi^2}\ln(s) + \frac{3\pi^2}{8N^2} + k_2 + k_3 N^2, \tag{2.2}$$

where

$$k_1 = \quad 2.135730074078457(2), \tag{2.3}$$

$$k_2 = -2.8626215972(6), \tag{2.4}$$

$$k_3 = \quad 1.24911585(3). \tag{2.5}$$

We do not comment on the size of $d_2(s)$ here, but would like to first put our result in the context of some concrete applications and shall then discuss the magnitude of the two-loop corrections in the relevant formulae.

3. As already mentioned in sect. 1, we are mainly interested in the expansion (1.1) relating the $\overline{\mathrm{MS}}$ coupling to the SF coupling. By combining our results with those of Narayanan and Wolff [19,20], we can now calculate $c_1$ and $c_2$ for $N = 2$ and obtain

$$c_1(s) = -\frac{11}{3\pi}\ln(s) + 0.94327(4), \tag{3.1}$$

$$c_2(s) = c_1(s)^2 - \frac{17}{3\pi^2}\ln(s) + 0.5216(5). \tag{3.2}$$

For $N = 3$ the expansion of $\alpha_{\rm SF}$ in powers of the bare coupling is currently only known to one-loop order so that in this section attention will be restricted to the SU(2) theory.

Before making practical use of the series (1.1) the scale factor $s$ should be fixed so that the low-order terms are reasonably small (as far as possible). If we choose

$$s = \Lambda_{\overline{\rm MS}}/\Lambda_{\rm SF} = 2.24385(8), \tag{3.3}$$

one gets $c_1(s) = 0$ and $c_2(s) = 0.0576(5)$, i.e. in this case the known perturbative corrections are indeed very small. Note incidentally that choosing $s$ to be equal to the appropriate ratio of $\Lambda$–parameters proved to be successful in other instances as well, where the applicability of the perturbation expansion could be controlled through numerical simulations [8].

The SF coupling in the SU(2) theory has been accurately determined in ref.[8] over a large range of momenta given in units of Sommer's scale $r_0$ [30]. Using the two-loop series (1.1) with $s$ as given above, these results may now be converted to the $\overline{\rm MS}$ scheme. At $q = 200/r_0$ for example (which corresponds to about 80 GeV in physical units), one obtains

$$\alpha_{\overline{\rm MS}}(q) = 0.1289(15)(3). \tag{3.4}$$

The first error here is the total statistical error as quoted in ref.[8], while the second is equal to $\alpha_{\rm SF}(q/s)^4$ and thus indicates the expected order of magnitude of the three-loop correction in eq.(1.1). Since the two-loop term is anomalously small, it has practically no influence on the central value of $\alpha_{\overline{\rm MS}}(q)$. The main impact of our computation of the two-loop correction is that the estimated error from the neglected higher order terms has been reduced from a few percent to a fraction of a percent.

A remarkable property of the SF coupling is that its evolution in the momentum range covered by the numerical simulations is accurately described by the two-loop approximation to the Callan-Symanzik $\beta$–function [3,8]. A perfect fit of the data was in fact obtained by adding an effective three-loop term with coefficient

$$b_2^{\rm eff} = 0.35(12)/(4\pi)^3. \tag{3.5}$$

(cf. appendix A for unexplained notation). From the above and the known value of $b_2^{\overline{\rm MS}}$ [31] we may now calculate the exact coefficient

$$b_2^{\rm SF} = 0.1797(3)/(4\pi)^3. \tag{3.6}$$

There is no reason to expect that $b_2^{\rm SF}$ is equal to $b_2^{\rm eff}$, but it is reassuring to see that both coefficients have the same sign and order of magnitude. Eq.(3.6) is certainly compatible with the numerical data.

4. More direct ways of using the bare perturbation expansion (1.2) have recently been discussed in refs.[10,32]. The two-loop corrections in such applications may now be calculated and we here consider two cases to illustrate this. For simplicity we set $N = 3$ throughout this section.

If we choose

$$s = \Lambda_{\overline{\rm MS}}/\Lambda_{\rm lat} = 28.80934(1) \tag{4.1}$$

(as suggested by previous experience), the one-loop term in eq.(1.2) vanishes and $d_2(s) = 4.44915(1)$. One might think that with a different scale factor the rather large two-loop coefficient can perhaps be reduced significantly. But this is not the case, since $d_2(s) \geq 4.31303(1)$ for all $s$.

In view of the size of the two-loop correction and the large scale factor (4.1) one may be hesitating to make practical use of this form of the expansion. The problem with the scale factor disappears if instead of $\alpha_0$ one uses the "improved" bare coupling [25,32]

$$\tilde{\alpha}_0 = \alpha_0/P \tag{4.2}$$

as the expansion parameter. To two-loop order the (normalized) plaquette expectation value $P$ appearing in this definition is given by [33,34]

$$P = 1 - \frac{4\pi}{3}\alpha_0 + \left(8k_2 + \frac{16\pi^2}{9}\right)\alpha_0^2 + \ldots, \tag{4.3}$$

where $k_2$ is the constant (2.4) introduced earlier. By combining eq.(1.2) with eqs.(4.2),(4.3) the desired expansion of $\alpha_{\overline{\rm MS}}(s/a)$ in powers of $\tilde{\alpha}_0$ is obtained straightforwardly. We again choose the scale factor so that the one-loop term vanishes and end up with

$$\alpha_{\overline{\mathrm{MS}}}(s/a) = \tilde{\alpha}_0 + 2.18505(1) \times \tilde{\alpha}_0^3 + \ldots, \qquad s = 2.63285(1). \tag{4.4}$$

This expansion looks indeed more comfortable although the two-loop coefficient is still rather large. In the relevant range of bare couplings, $g_0^2 \simeq 1$, the correction is 3–4%.

The introduction of the "improved" bare coupling (4.2) has partly been motivated by the observation that the dominant contributions to the one-loop coefficient $d_1(s)$ come from tadpole diagrams and that these cancel when $\alpha_0$ is replaced by $\tilde{\alpha}_0$. Assuming tadpole dominance at two-loop order, as proposed by Ellis and Martinelli [22–24], however results in an estimate of the coefficient quoted in eq.(4.4) which is too small by more than a factor 3. An uncritical application of tadpole dominance can thus be rather misleading.

5. The extension of our calculations to lattice QCD with Wilson quarks is straightforward. We also expect that the expansion of $\alpha_{\mathrm{SF}}$ in powers of the bare lattice coupling can be worked out for this case to two-loop order along the lines explained in ref.[20]. Such computations would certainly be welcome, but it may take a while until the precision on $\alpha_{\mathrm{SF}}$ which can be reached in numerical simulations of full QCD is comparable to the expected order of magnitude of the two-loop correction in eq.(1.1).

We would like to thank Claus Vohwinkel for helpful discussions and Anton van de Ven for performing some additional checks on our computations.

A. For any running coupling $\alpha = \bar{g}^2/4\pi$ depending on a momentum scale $q$ we define the associated $\beta$–function through

$$\beta(\bar{g}) = q\frac{\partial \bar{g}}{\partial q}. \tag{A.1}$$

In perturbation theory we have

$$\beta(\bar{g}) = -\bar{g}^3 \sum_{n=0}^{\infty} b_n \bar{g}^{2n} \tag{A.2}$$

with $b_0 = 11N/48\pi^2$ and $b_1 = 17N^2/384\pi^4$ being the universal coefficients. All other (three-loop and higher order) coefficients depend on the chosen scheme.

The $\Lambda$–parameter is defined by

$$\Lambda = \lim_{q\to\infty} q\,(b_0\bar{g}^2)^{-b_1/2b_0^2}\mathrm{e}^{-1/2b_0\bar{g}^2}, \tag{A.3}$$

and if

$$\bar{g}'^{\,2} = \bar{g}^2 + r_1\bar{g}^4 + r_2\bar{g}^6 + \ldots \tag{A.4}$$

is another running coupling, we have

$$\Lambda' = \Lambda\mathrm{e}^{r_1/2b_0}, \tag{A.5}$$

$$b_2' = b_2 - b_1 r_1 + b_0(r_2 - r_1^2). \tag{A.6}$$